\documentclass[prb,aps,twocolumn,superscriptaddress]{revtex4-2}
\usepackage{amsmath}
\usepackage{amsfonts}
\usepackage{amssymb}
\usepackage{graphicx}
\usepackage{mathtools}
\usepackage{array}
\usepackage[colorlinks=true,citecolor=blue,linkcolor=blue]{hyperref}
\usepackage{mathrsfs}
\usepackage{xcolor}
\definecolor{XQ}{rgb}{1,0,0}

\newcommand*{\RuCl}{\ensuremath{{\alpha\mbox{-}\mathrm{RuCl}_3}}}
\newcommand*{\K}{\ensuremath{{\mathrm{K}}}} 
\newcommand*{\T}{\ensuremath{{\mathrm{T}}}} 
\makeatletter   

\newcommand*\wthelper[2]{%
    \hbox{\dimen@\accentfontxheight#1%
        \accentfontxheight#11.15\dimen@
        $\m@th#1\widetilde{#2}$%
        \accentfontxheight#1\dimen@
    }%
}
\newcommand*\accentfontxheight[1]{%
    \fontdimen5\ifx#1\displaystyle
        \textfont
    \else\ifx#1\textstyle
        \textfont
    \else\ifx#1\scriptstyle
        \scriptfont
    \else
        \scriptscriptfont
    \fi\fi\fi3
}
\makeatother

\begin{document}

\author{Zheng-Duo~Fan}
\affiliation{Institute for Advanced Study, Tsinghua University, Beijing, 100084, China}
\author{Xiao-Qi~Sun}
\affiliation{Department of Physics and Institute for Condensed Matter Theory, University of Illinois at Urbana-Champaign, Urbana, IL 61801, USA}
\affiliation{Max-Planck-Institut f{\"u}r Quantenoptik, Hans-Kopfermann-Stra{\ss}e 1, D-85748 Garching, Germany}
\affiliation{Munich Center for Quantum Science and Technology (MCQST), Schellingstra{\ss}e 4, D-80799 M{\"u}nchen, Germany}
\author{Jing-Yuan~Chen}
\affiliation{Institute for Advanced Study, Tsinghua University, Beijing, 100084, China}

\title{On the Thermal Transport Puzzles in $\RuCl$}
\date{\today}

\begin{abstract}
Thermal transport has been used to probe the nature of $\RuCl$, an important candidate of Kitaev material. Two remarkable observations were made under applied magnetic fields at low temperatures, and have stimulated extensive discussions. One is a sizable thermal Hall effect, and the other is an apparent ``oscillation'' of the longitudinal thermal conductivity with the magnetic field. It has been proposed that the former is due to a bosonic Chern band. Meanwhile, the origin of the latter has largely remained obscure. This work aims to resolve this ``oscillation'' puzzle. By examining the thermal transport data as well as other measured properties of $\RuCl$, we argue that the most plausible scenario is that of phonons scattering with spin degrees of freedom across multiple phases. We substantiate this picture into a phenomenological theory, which reproduces the ``oscillation'' behavior in a simple manner and makes predictions that can be examined by future experiments. Moreover, our phenomenological theory and the aforementioned proposal for the thermal Hall effect support each other. We hope this work can thus help settle the physical mechanism behind the thermal transport puzzles in $\RuCl$.
\end{abstract}

\maketitle

\tableofcontents

\section{Introduction}
\label{sect_intro}

Thermal properties often reveal important aspects of a condensed matter system. The rough idea is, the system may contain interesting degrees of freedom, fundamental or emergent, which do not carry electric charge; however they do always carry energy, leading to signatures in thermal properties. A recent remarkable success in this regard is the observation of half-quantized thermal Hall conductance in the $\nu=5/2$ quantum Hall state \cite{5/2quantumhall}, which revealed the existence of chiral Majorana edge mode and hence, for the first time, nailed down the observation of a non-abelian topological order. For a similar motivation, in the recent years thermal transport has been used extensively in the hunting for materials realizing Kitaev spin liquid, another non-abelian topological order which also hosts a chiral Majorana edge mode, as has been theorized in Ref.~\cite{kitaevspinliquid}. Among its candidate materials, one that is proposed the earliest and attracted wide interests is $\RuCl$, see e.g. \cite{spinliquid1,spinliquid2,spinliquid3,spinliquid4} for reviews. And the recent thermal transport measurements on $\RuCl$ have indeed brought in surprises and generated extensive discussions.

Our goal of this paper is to carefully analyze the thermal transport data in $\RuCl$ and, combining with its other measured properties, figure out what these data have revealed to us about this material. 
We will layout our general reasoning, introduce our phenomenological theory and make predictions to be examined by future experiments.

\begin{figure}
    \centering
    \includegraphics[width=0.49\textwidth]{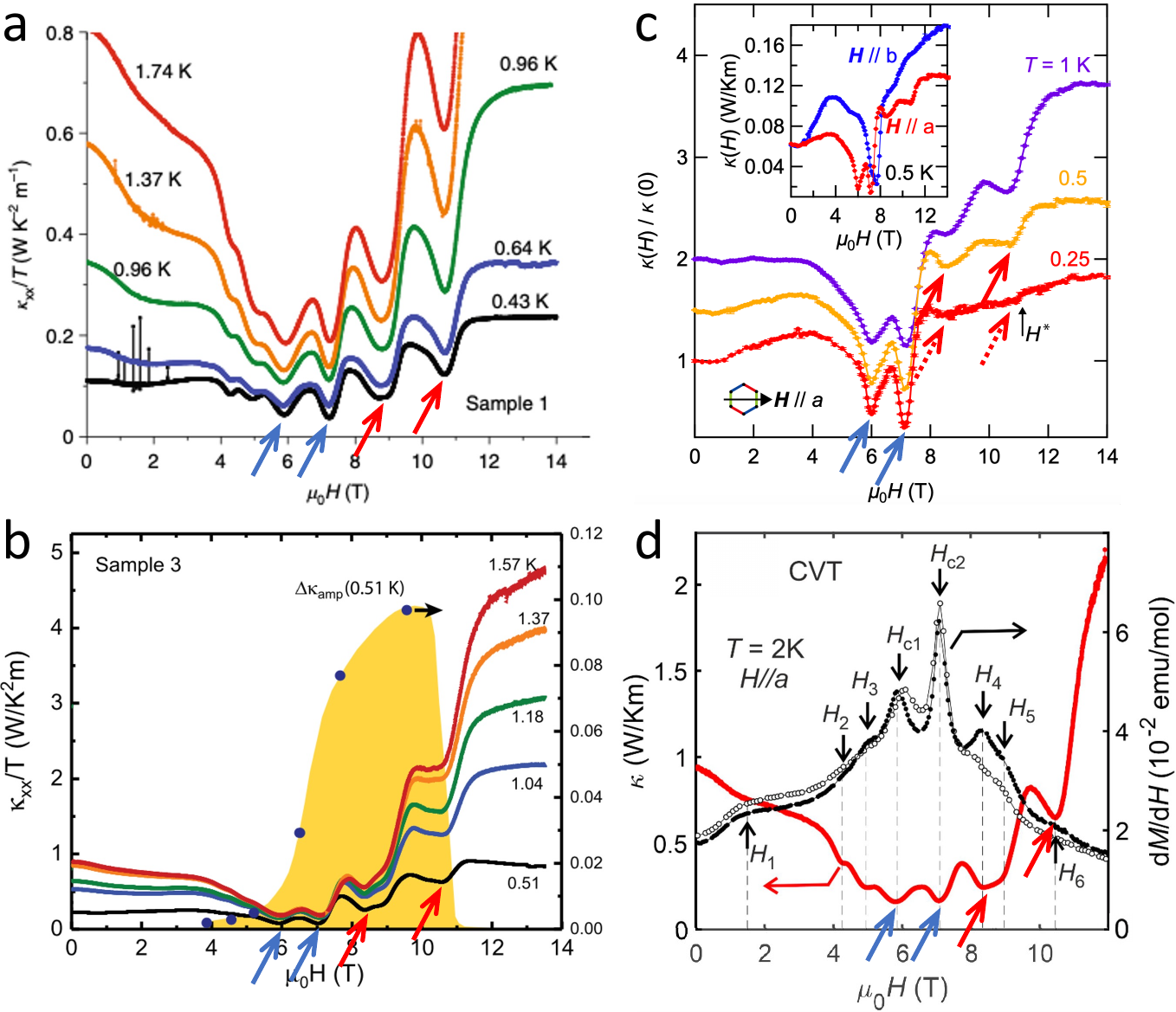}
    \caption{Longitudinal thermal conductivity of \RuCl. (a) and (b) are both reprinted from \cite{Ong1}, and they come from different samples, (c) is reprinted from \cite{Matsuda1}, and (d) from \cite{Takagi1}. We added the tilted arrows indicating the two sets of dips (see Section \ref{ssect_twosets}) that we will investigate. The blue arrows indicate the first set of dips, and the red  arrows the second set. The dashed red arrows in (c) emphasize the disappearance of the second set of dips at the lowest temperature so far observed, which is an important feature that we will explain.}
    \label{fig:expt_summary_long}
\end{figure}

$\RuCl$ is a layered, quasi-2D material. Its basic properties can be found in e.g. Refs.~\cite{abinitio,heatcapacity,magneticorder}. The thermal transport observations of interest, made at low temperatures and under applied in-plane magnetic fields, can be summarized as two puzzles:
\begin{enumerate}
    \item A sizable thermal Hall conductivity $\kappa_H$ of order $1$ magnitude in its natural unit $\kappa_0=\pi^2 k_B^2 T/6\hbar$ has been observed \cite{Matsuda2,Takagi2, Ong2}.

    \item The longitudinal thermal conductivity $\kappa_L$ exhibits significant ``oscillations'' with the in-plane magnetic field at lower temperatures \cite{Ong1}. In Fig.~\ref{fig:expt_summary_long} we present the results from Refs.~\cite{Ong1}, \cite{Matsuda1} and \cite{Takagi1}.
\end{enumerate}
Since its first observation, the thermal Hall conductivity has been speculated to be carried by the long sought chiral Majorana edge mode in Kitaev spin liquid \cite{Matsuda2}; however, the more recent data \cite{Ong2} strongly suggests that the heat carrier has more of a Bose-Einstein statistics character. One plausible phenomenological explanation is topological magnon bands \cite{Kimmagnon}, as we will discuss in the main text later. On the other hand, the ``oscillations'' of the longitudinal thermal conductivity has insofar remained even more obscure. Therefore the primary problem of this paper is to explain this second puzzle, in hope that along the way the explanations for these two puzzles can provide support for each other.

Let us summarize our phenomenological conclusions here, before presenting the details of our analysis of the experimental data and our theoretical reasoning. 
\begin{enumerate}
    \item The ``oscillating'' behavior is highly unlikely to be a quantum oscillation of any sort -- by which we mean \emph{any} mechanism due to the $U(1)$ nature of the magnetic field, or say the $2\pi(\hbar/e)$ periodicity of the magnetic flux.
    \item The longitudinal thermal current is carried by acoustic phonons as usual. The ``oscillating'' behavior is due to a series of magnetic field driven phase transitions of the spin degrees of freedom in the material, which results in collective degrees of freedom that scatter the phonons strongly. In particular, the scatterers responsible for some dips in $\kappa_L$ are the critical fluctuations at second order quantum phase transitions, while those responsible for other dips may be domain walls formed near first order quantum phase transitions.
    \item With some physical insights from the previous studies on $\kappa_H$, we can form simple phenomenological models for these two kinds of mechanisms, which are consistent with other measured properties of the material. Moreover, we make predictions that can be examined by future experiments.
\end{enumerate}  
From the currently available data on the thermal properties of $\RuCl$ and our analysis, it is premature to conclude on the precise nature of those underlying phases, especially whether or not there is really a spin liquid phase. However, our conclusions do provide useful constraints and insights on what the phases might do.

This paper is organized as the following. In Sect.~\ref{sect_general} we present our general considerations which settle the direction of thinking. In Sect.~\ref{sect_firstset} and \ref{sect_secondset} we introduce two mechanism responsible of two sets of dips in $\kappa_L$ respectively. In Sect.~\ref{sect_Hall} we explain how our theory and the topological magnon theory for thermal Hall conductivity \cite{Kimmagnon} support each other. In Sect.~\ref{sect_predictions} we summarize our results and list the predictions for future experiments to examine our phenomenological picture.

\section{General Considerations}
\label{sect_general}

Our analysis is data driven, which means \emph{a priori} we are open to different directions of theoretical modeling. It is therefore important to first narrow down the directions using the data, numerology and qualitative reasoning.

\subsection{Difficulties with Quantum Oscillation}
\label{ssect_qo}

The first question of importance, in regard of the observed ``oscillating'' behavior in $\kappa_L$, is whether it is a quantum oscillation of any sort. If it is, since $\RuCl$ is definitely not a Fermi liquid of electrons, it must be of highly exotic nature. Now we argue that any quantum oscillation is an unlikely scenario.

By ``quantum oscillation'', we broadly mean \emph{any} oscillation mechanism that has an origin in the $U(1)$ nature of the magnetic field, i.e. the $2\pi$ periodic nature of the magnetic flux in the natural unit $\hbar/e$. In this very broad sense, the oscillation can conceivably take periods either in $1/B$ (for instance the usual quantum oscillation in an electronic Fermi liquid, where something special happens when on average an area of $2\pi$ flux contains an integer number of mobile charges) or in $B$ (if something special happens when the flux through some fixed area crosses an integer multiple of $2\pi$).

In the $\kappa_L$ measurements of $\RuCl$, the magnetic fields are at the order of $B\sim 10\T$, and in this range only a few dips are observed. This means, if the phenomenon is indeed some form of quantum oscillation, regardless of whether it takes periods in $1/B$ or $B$, the characteristic magnetic field must be of order $\sim 10\T$. The characteristic flux per unit cell area is $B a^2\sim 5\times 10^{-3}$ in the natural unit $\hbar/e$. This is $10^{-3}$ too small compared to the $2\pi$ needed for quantum oscillation. This means a large compensation factor of $10^3$ is needed somewhere. In the case of electronic Fermi liquid, the integer quantum Hall effect as a limiting case of quantum oscillation also has a comparable characteristic magnetic field. There, the large $10^3$ factor needed comes from the fact that the average area occupied by each mobile electron is much larger than a unit cell -- but this is due to the artificial fine-tuning of the 2DEG density to be so low for this very purpose. By contrast, in the insulator $\RuCl$ there does not appear to be any degree of freedom associated with such a large area. Any strong assumption that places back the missing numerical factor of $10^3$ would be highly contrived, regardless of the details of the theoretical proposal (think of some emergent magnetic field in place of the applied magnetic field, but somehow $10^3$ larger, or some emergent mobile charged degrees of freedom that are present but somehow as dilute as $10^{-3}$ per unit cell).

To add to the unlikeliness of quantum oscillation are two more specific considerations: 1) in these measurements the magnetic field is in-plane rather than perpendicular, and 2) the obvious degrees of freedom coupling to the magnetic field in the material should be the spins, which do not see the $2\pi$ periodic nature of magnetic flux to begin with.

The arguments above are from the theoretical side. From the data side there is no compelling evidence for a $1/B$-periodic quantum oscillation either (and the ``oscillation'' is obviously not periodic in $B$). In Sample 1 of \cite{Ong1}, six dips were observed, such that the first three fits to one value of periodicity in $1/B$ while the latter three fits to another value which differs by nearly 50\%; in the chemical vapor transport (CVT) sample of \cite{Takagi1}, a similar six dips pattern was observed. On the other hand, in Sample 3 of \cite{Ong1}, in the Bridgman sample of \cite{Takagi1}, and in the samples in \cite{Matsuda1}, only four dips were observed. So from the currently available data combined, it is hard to conclude a $1/B$-periodic behavior.

\subsection{Why Phonon Scattering}
\label{ssect_phonon}

How do we perceive the unusual signature in the thermal conductivity, if not as any kind of quantum oscillation? We argue that it is most plausible to view the signature as a series of \emph{dips} occurring in the thermal conductivity, as a result of the heat-carrying acoustic phonons being scattered by collective degrees of freedoms formed by the spins.

$\RuCl$ is, in the first place, being studied for its rich spin dynamics. \emph{A priori}, the spin dynamics is expected to be complex and hard to analyze, therefore a good starting point is to first look at the large magnetic field limit in which the spin degrees of freedom are frozen to align with the field. We note in Fig.\ref{fig:expt_summary_long} that:
\begin{itemize}
    \item For $B> 12\T$, the value of $\kappa$ stabilizes, and notably this value is larger than the values at any lower magnetic fields, including the ``oscillation'' region $4\T < B < 12\T$.
\end{itemize}
(Here and in the below, we will use the bullet point $\bullet$ for aspects of the experimental results that we want to emphasize.) In this large $B$ limit where the spin dynamics is frozen, the $\kappa$ value should entirely come from the usual acoustic phonon contribution; it is found in this limit that $\kappa$ exhibits the phonon-driven $T^3$ behaviour \cite{Takagi2}. The fact that $\kappa$ is lower at smaller magnetic fields strongly suggests that, as the spins become more dynamical, they scatter off phonons, and the dips are where the spins form particularly strong scatterers.

We can further constrain the properties of such scatterers. At low temperatures, the excited acoustic phonons are of low energy and long wavelength, and therefore usually interact weakly with other degrees of freedom -- the acoustic wave often simply diffract through, as is familiar from classical physics. With this in mind, it becomes clear that: 
In order for the acoustic phonons to be scattered strongly, the scatterer must either has low energy dynamics, so that resonance occurs, or has large size, so that the long wavelength acoustic wave sees it easily rather than diffracts through.
Therefore, the plausible physical picture is, the spins have rich dynamics among themselves, such that at certain magnetic fields, collective low energy and/or long wavelength degrees of freedom are formed, and they interact strongly with the acoustic phonons, hence turning the magnetic field driven effects into dips in the thermal conductivity. (As we will see later, this picture is also consistent with the explanation to the sizable thermal Hall effect.) Note that the spin system itself may also carry some heat current, but the formation of dips means that such positive contribution is minor compared to the negative contribution due to the strong scattering.

Before we move on, we would like to remark that the picture of spin-phonon interaction gives reasonable numerology (as opposed to quantum oscillation). Spin couples to the magnetic field through energetics. The magnetic dipole moment of a spin is of order $1$ Bohr magneton, which in the relevant units is $\mu_B\simeq 0.7 \K/\T$. Therefore, in the typical magnetic field $B\sim 10\T$ in these experiments, the characteristic energy scale is $B\mu_B \sim 7\K$, which is at the same order as the temperature which excites the acoustic phonons.

\subsection{Two Sets of Dips}
\label{ssect_twosets}

Now let us examine those unusual features more closely. We note the following:
\begin{itemize}
    \item As we mentioned at the end of Sect.~\ref{ssect_qo}, there are four dips that are observed in the all samples prepared by different methods and measured by different groups \cite{Ong1,Matsuda1,Takagi1}. We will focus on these dips. They occur at $B\approx 6\T, 7\T, 9\T, 11\T$ respectively. In Ref.~\cite{Matsuda1} and Ref.~\cite{Takagi1}, the measurements are made down to $0.15\K$, and it is found that the two dips at $B\approx 6\T$ and $7\T$ persist at these low temperatures, whilst the other two at $B\approx 9\T$ and $11\T$ fade away below $0.25\K$.
    \item The phase diagram of $\RuCl$ in magnetic field and temperature has been consistently settled using different approaches, including heat capacity, susceptibility, neutron scattering, etc. \cite{heatcapacity,magneticorder}. The interval of $B$ in which the dips in $\kappa$ occur runs cross multiple phases of the spin system: two zig-zag magnetic ordered phases and the debated ``quantum spin liquid'' regime which is not well-understood.
\end{itemize}
These observations suggest that the ``oscillation'' might be caused by different mechanisms in different phases of the spin system, instead of a single mechanism. It is reasonable to separate the dips into two sets which behave differently at low temperatures.

For the first set of dips at $B\approx 6\T$ and $7\T$,
\begin{itemize}
    \item It is noted in Ref.~\cite{Takagi1} that these two dips in $\kappa$ coincide with two peaks in magnetic susceptibility, which correspond to the two second order magnetic phase transitions: $B_{c1}$ between the two zig-zag phases, and $B_{c2}$ from a zig-zag phase to the debated ``quantum spin liquid'' regime.
\end{itemize}
This strongly suggests that the scatterers for these two dips are the low energy, long wavelength critical fluctuations near the transitions. In Section \ref{sect_firstset}, we will formulate this physical picture with the theory of critical scattering.

The second set of dips at $B\approx 9\T$ and $11\T$ is more mysterious. They occur in the not-well-understood regime in the phase diagram and are not accompanied by other apparent features. In particular, heat capacity has been measured in this regime of magnetic fields \cite{heatcapacity} and the temperature scaling is dominated by the $T^3$ from acoustic phonons, up to exponentially small corrections. This shows the absence of low energy spin excitation modes below $\sim 10\K$. By the reasoning from the previous subsection, this means the scatterers are likely large size extended objects. In \cite{Takagi1} it was suggested that the scatterers are stacking faults. However, the temperature dependence of the magnetic susceptibility in \cite{Ong1} shows the absence of stacking fault (which is, in particular, in contrast to the sample in \cite{Takagi1} prepared by the same CVT method); across all the samples currently available, there is no clear correlation between the prominence of the dips and the amount of stacking faults. Instead, we will show in Section \ref{sect_secondset} that a likely scenario is that the scatterers are domain walls between competing spin states in this regime, in consistency with the explanation for the thermal Hall measurements (see Section \ref{sect_Hall}).

\section{First Set of Dips}
\label{sect_firstset}

In this section we will discuss the two dips near the critical points $B_{c1}$ and $B_{c2}$, due to phonons scattering with the long wavelength, low energy critical fluctuations. Historically such effect has been studied by the theory of \emph{critical scattering}, which first appeared in the context of critical opalescence, and later also in the context of N\'{e}el phase transition \cite{Stern}.  However, the original derivation of the theory was somewhat non-transparent and moreover contained some important caveat. In this section we will introduce the theory in a more accessible form and cure the caveat.

The key idea and the beauty of this theory is that the effects can be studied without involving much details of the spin system -- the phonon dynamics is much slower than the spin dynamics, hence only sees the hydrodynamical behavior of the spin system, and hydrodynamics is universal. 

Consider a generic spin Hamiltonian
\begin{eqnarray}
H_{s}=\sum_i \mathcal{H}_i=\sum_{i, \alpha} J_\alpha Q_i^\alpha \ ,
\end{eqnarray}
where $i$ labels the lattice site, and $\alpha$ labels different spin interaction terms in the Hamiltonian; in particular, $Q_i^\alpha$ is a local spin interaction term near site $i$, and $J^\alpha$ is the corresponding coupling constant. For instance, in $\RuCl$ it is believed that the important contributions have $\alpha$ running over Heisenberg terms and Kitaev terms in different link directions; we emphasize that these details are unimportant for the critical scattering.

Phonons and spins are coupled through the dependence of the spin coupling constants $J^\alpha(\mathbf{u})$ on the variation of the acoustic displacement $\mathbf{u}$. Schematically we can write the interaction as
\begin{eqnarray}
V= \sum_{i,\alpha} \left[ J_\alpha^{(1)} (\partial u)_i + J_\alpha^{(2)} (\partial u)_i^2 + \cdots \right] Q_i^\alpha
\label{interactions}
\end{eqnarray}
In this expansion we kept the $\partial u$ terms which correspond to the spin system absorbing or emitting a phonon, and then the $(\partial u)^2$ terms which correspond to the spin system absorbing and then emitting a phonon, or absorbing or emitting two phonons.
As the spin dynamics are in general much faster than the phonon dynamics, resonance is hard to occur meanwhile conserving momentum and energy, therefore resonance is not a major enhancement mechanism. On the other hand, processes that are elastic in energy but involving large momentum exchange between the phonons an the spin system can take advantage of long wavelength of the critical fluctuations in the spin system. Such elastic processes are the processes of absorbing and then emitting a phonon, which appear in the $(\partial u)^2$ terms. (This argument can be verified more quantitatively \cite{Stern}, as it will become clear later.) In the following, we therefore only consider the $(\partial u)^2$ terms, and we denote them as $\sum_i (\partial u)_i^2 \mathcal{V}^{(2)}_i$, where $\mathcal{V}^{(2)}_i= \sum_\alpha J_\alpha^{(2)} Q_i^\alpha$.

We will describe the phonon transport using the Boltzmann equation, and the scattering effects occur in the collision term. It is easy to see that the elastic scattering contribution to the relaxation rate of phonons is schematically proportional to:
\begin{eqnarray}
&& 1/\tau({\bf k}) \propto \int_{\bf k'} A_{\mathcal{V}^{(2)}}(\omega_{\bf k'}-\omega_{\bf k}, {\bf k'-k }) \ , \\
&& A_{\mathcal{V}^{(2)}}(\omega, \mathbf{q}) \equiv \sum_{i} \int_{-\infty}^\infty dt \ e^{i\mathbf{q}\cdot (\mathbf{r}_i-\mathbf{r}_{i_0}) - i\omega t} \nonumber \\ 
&& \hspace{3cm} \left\langle \mathcal{V}^{(2)}_i(t) \: \mathcal{V}^{(2)}_{i_0}(0) \right\rangle_s  \nonumber
\end{eqnarray}
where $\omega_{\bf k}$ is the energy of an acoustic phonon of momentum $\bf{k}$, the expectation $\langle \cdots \rangle_s$ is taken in the spin system, and the $t$ dependence on $\mathcal{V}^{(2)}$ is generated by the spin dynamics $H_s$.

The correlation $A$ is not in the usual (and physical) retarded ordering. However, it can be related to the retarded one by the fluctuation-dissipation theorem (which can be proven by the spectral representation):
\begin{eqnarray}
A_{O}(\omega, \mathbf{q}) = -\frac{\mathrm{Im}\chi_O(\omega, \mathbf{q})}{1-e^{-\omega/T}} 
\end{eqnarray}
where $\chi_O(\omega, \mathbf{q})$ is the Fourier transformation of the retarded correlation between local Hermitian operators $O$. Even so, the correlation in the spin system is still formidable to compute directly, because the spin system of interest is strongly interacting. Fortunately, in our case, since the phonon dynamics is much slower than the spin dynamics, we only need to consider the slow response in the spin system, which should be universally described by hydrodynamics near the gapless criticality. In hydrodynamics, the only remaining correlations are those between conserved current components. While $\mathcal{V}^{(2)}_i$ is itself not a conserved current component, we note that it involves the same operators $O^\alpha_i$ with the Hamiltonian density $\mathcal{H}_i$, though with somewhat different coefficients. Therefore, we may replace $\mathcal{V}^{(2)}_i = a \mathcal{H}_i + \cdots$, where the coefficient $a$ is non-small, and the remaining terms are unimportant and dropped for not being conserved current components. Therefore, in summary, we have
\begin{eqnarray}
1/\tau({\bf k}) \propto A_{\mathcal{H}}(\omega_{\bf k'}-\omega_{\bf k}, {\bf k'-k }),
\end{eqnarray}
where $A_\mathcal{H}$ is proportional to the retarded correlation $\chi_{\mathcal{H}}$ between spin Hamiltonian densities, a hydrodynamical quantity.

The main conclusion of critical scattering, given the picture formulated above, is that $1/\tau$ reaches maximum with the heat capacity of the spin system -- hence at the criticality. The heat capacity $C_s$ is proportional to $\mathrm{Im}\chi_{\mathcal{H}}(\omega, \mathbf{q})$ in the $\omega/|\mathbf{q}|\rightarrow 0, |\mathbf{q}|\rightarrow 0$ limit, so what we have to do is to analyze the $\omega, \mathbf{q}$ dependence of $\mathrm{Im}\chi_{\mathcal{H}}$. In Appendix \ref{app_crit_scatt}, we show that in hydrodynamics,
\begin{eqnarray}
    \mathrm{Im}\chi_{\mathcal{H}}(\omega, \mathbf{q}) = \frac{\omega D_s q^2}{(D_s q^2)^2+\omega^2} \: \chi_{\mathcal{H}}(0, \mathbf{q})
    \label{ImChi_factors}
\end{eqnarray}
where $D_s$ is the heat diffusion coefficient of the spin system. 

Let us first look at the $\chi_H(0, q)$ factor in \eqref{ImChi_factors}. From critical scaling, we know that as the spin correlation length $\xi$ diverges,
\begin{eqnarray}
    C_s\propto\chi_H(0, 0) \xrightarrow{\xi\rightarrow \infty} \xi^{\alpha/\nu}, \nonumber \\
    \mbox{and } \chi_H(0, q) \xrightarrow{\xi\rightarrow \infty} q^{-\alpha/\nu},
\end{eqnarray}
where $\alpha, \nu$ are the familiar critical exponents. This suggests we can write $\chi_{\mathcal{H}}(0, q) \sim q^{-\alpha/\nu} f(q\xi)$, where $f(x)$ is some function satisfying $f(x\rightarrow 0)\rightarrow x^{\alpha/\nu}$ and $f(x\rightarrow \infty)\rightarrow 1$. If we further make the mild assumption that $f(x)$ is monotonic, then we can see $\chi_{\mathcal{H}}(0, q)$ at fixed finite $q$ increases with $\xi$, though not dramatically. (In \cite{Stern}, the $q$ in $\chi_{\mathcal{H}}(0, q)$ was taken to $0$ as an approximation without justification. Our discussion here clarifies this caveat.)

Next let us look at the remaining, $\omega$ dependent factor in \eqref{ImChi_factors}. We note that $A_\mathcal{H}\sim (T/\omega)\mathrm{Im}\chi_\mathcal{H}$ peaks at $\omega=0$. This peaking is particularly prominent near criticality, because the thermal conductivity of the spin system (note, \emph{not} the full system) is $\kappa_s\sim C_s D_s$, and at criticality, $C_s$ diverges (or is cutoff by some disorder scale) while $\kappa_s$ does not in normal cases, which means $D_s \sim \kappa_s/C_s$ becomes particularly small. Recall that here $\omega$ corresponds to $\omega_{\bf k'}-\omega_{\bf k}$, which can vanish for elastic scattering of phonons at finite density of states. (This, in retrospect, justifies our initial assumption that the $\mathcal{V}^{(2)}$ contribution dominates over the $\mathcal{V}^{(1)}$ contribution \cite{Stern}, because for the latter, the same formula applies but instead with $\omega$ corresponding to $\omega_{\mathbf{k}}$ of a phonon, and phonons of vanishing $\omega_{\mathbf{k}}$ have vanishing density of states.)

These arguments explain why the spins at criticality scatter phonons particularly strongly, leading to dips in the phonon thermal conductivity.

\section{Second Set of Dips}
\label{sect_secondset}

The origin of the two dips in the debated quantum spin liquid regime is more obscure. Since the gap of spin system in this regime ($\sim 10\K$, as discussed Sect.~\ref{ssect_twosets}) is well above the temperature ($<4\K$) of the longitudinal thermal conductivity experiments, the scatterers cannot be the low energy excitation of the spin system. As there are no other obvious experimental signatures found in coincidence with these two dips, we have to start with some educated hypothesis about the nature of the scatterers, and then compare the theoretical analysis with the thermal measurements. The principle behind our hypothesis is that the scatterer must have some typical scale comparable to the phonons -- either of low energy or of extended size. Now we discuss these two possibilities.

At the low energy front, given we have already excluded the spin state excitations, the remaining plausible possibility would be local dynamical defects, as is inspired by the traditional glass studies \cite{defect1} and the recent thermal Hall studies \cite{defect2,defect3}. We consider dynamical defects whose energy levels depend on the magnetic field, so that at certain magnetic fields the defects are near resonance with the phonons, giving rise to sizable scattering effect. We find such model can indeed produce dips in the magnetic field; however, the positions of the dips will be temperature dependent, while in the $\RuCl$ experiments the temperature dependence is negligible. Therefore, while such mechanism might be of general interest in thermal transport, it does not apply to $\RuCl$, and we put the details of this theory in Appendix \ref{app_dyn_defect}.

At the extended size front, we propose the following physical picture, which is supported by the thermal Hall explanation and some numerics \cite{Kimmagnon}, as we will discuss in Sect.~\ref{sect_Hall}. The picture is that there are competing low energy spin states forming domains; at certain magnetic fields, the competing energies are so close that the domain walls get particularly dense. The characteristic length scale here is the typical domain size, comparable to or larger than the acoustic phonon wavelength, so that the long wavelength acoustic waves sees them rather than diffracts through. In Sect.~\ref{ssect_domains}, we show the thermal transport features predicted by this picture, under very modest assumptions, match well with the experimental results, including some seemingly puzzling details. Meanwhile, questions can be raised about our picture and assumptions -- a major one is, why the competing spin states did not lead to other signatures, e.g. in heat capacity? We will justify these issues in Sect.~\ref{ssect_heatC}, showing that our theory is consistent with all the currently available relevant measurements on $\RuCl$.

\subsection{Theory and Thermal Conductivity}
\label{ssect_domains}

\begin{figure}
    \centering
    \includegraphics[width=0.35\textwidth]{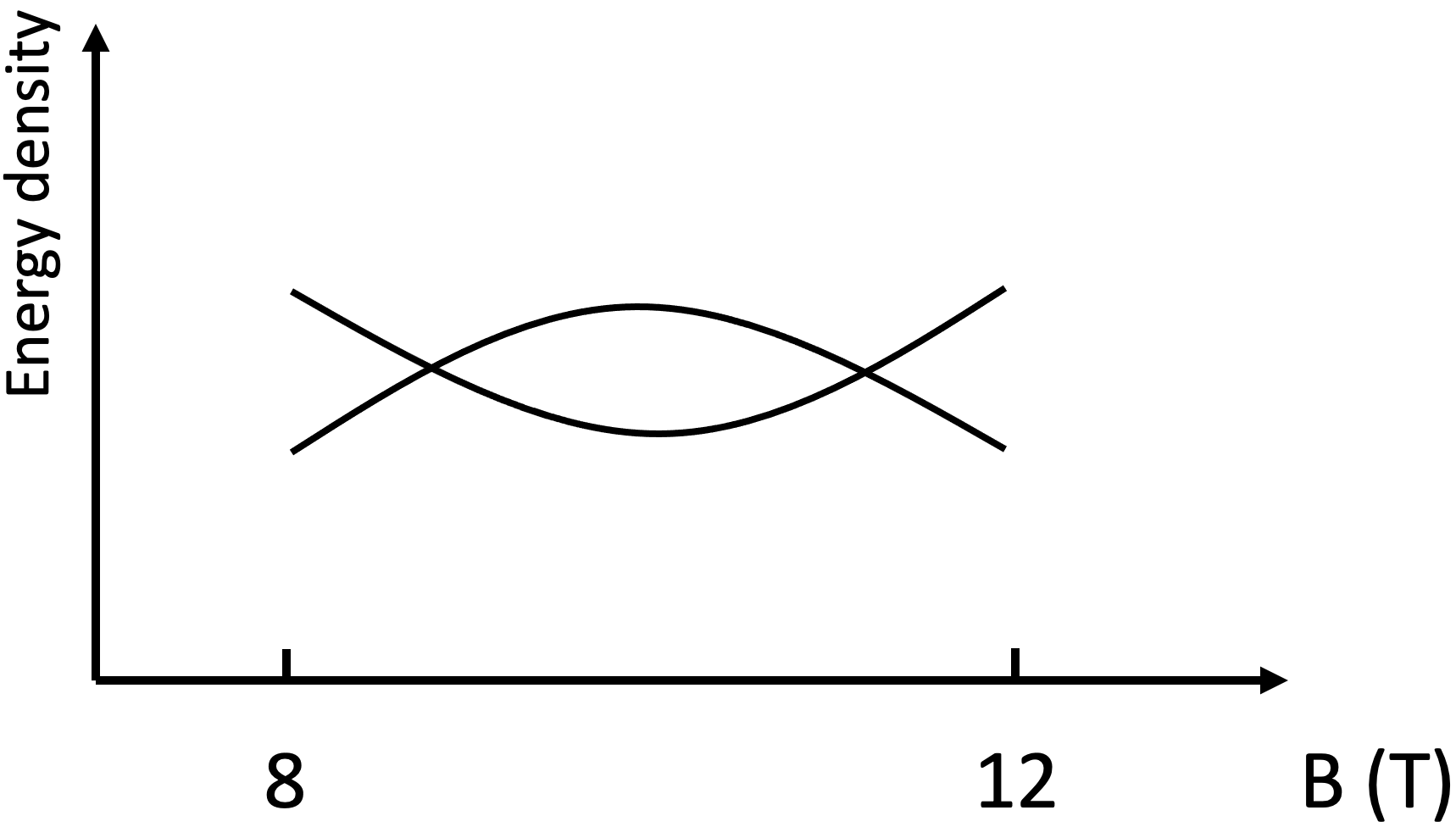}
    \caption{A schematic energy density competition between low energy spin states.}
    \label{fig:energy_crossings}
\end{figure}

This physical picture of domains of competing spin states is largely motivated by the interesting temperature dependence of $\kappa$ at and near the two dips of interest, see Fig.~\ref{fig:expt_summary_long}:
\begin{itemize}
    \item  The two dips of interest are observed for $0.5\K< T < 3\K$, and moreover the $\kappa$ values near the dips are significantly smaller than stable value at large $B$. For $T > 3\K$ \cite{Ong1, Takagi1, Matsuda1}, the two dips gradually disappear, while the overall $\kappa$ values at nearby magnetic fields remain significantly smaller than the stable value at $B> 12\T$. For $T<0.5\K$ \cite{Matsuda1,Takagi2}, the two dips disappear too, but now the overall $\kappa$ values at nearby magnetic fields are close to the stable value at $B> 12\T$.
\end{itemize}
These behaviors find simple interpretations in our picture. We assume the competing spin states have close energy densities following a pattern illustrated in Fig.~\ref{fig:energy_crossings}, in which the energy crossings are where the two dips of interest would occur. When the temperature is too high compared to the typical energy difference between the domains, both competing states appear from thermal fluctuations and form domains anyways, regardless of whether the energies cross at certain magnetic fields. Thus, the scattering of phonons from these domains leads to an overall reduced $\kappa$ compared to the $\kappa$ value at large $B$, but without obvious dips. On the other hand, when the temperature is too low, the state with lower energy will dominate over the sample, so there will be few domains and thus no obvious reduction of $\kappa$. (For this to happen, one might note that the spin state domains must be fairly easy to re-distribute in order to thermally re-equilibriate as the temperature varies in the experiments. This is in contrast to the familiar magnetic domains in permanent magnets, or structural domains, which freeze rather than re-equilibriate at low temperatures. We will justify this important point in Sect.~\ref{ssect_heatC} and Appendix~\ref{app_domain_dyn}.)

Now that we have this qualitative picture, we need to show that under moderate, detail-independent assumptions, this picture can indeed lead to the observed reduction of $\kappa$ -- for $T\approx 1\K$, the overall $\kappa$ values in the debated spin liquid regime $8\T < B< 12\T$ is about half of the stable value at large $B$. The domain walls can scatter the phonons in many ways. There are refraction and reflection, as long as phonons propagate at different sound speeds in different spin state domains, due to the spin-phonon interactions Eq.\eqref{interactions}; moreover, there may exist dynamical degree of freedom right on the domain wall with which the phonons interact. In the following, we only consider refraction and show this already gives us sufficient scattering under mild assumptions.

We will assume the sounds speeds in the two competing spin states to have a ratio of order
\begin{eqnarray}
v'/v\sim 0.95.
\end{eqnarray}
While we do not need to make detailed assumptions about the nature of the competing spin states, this ratio is a reasonable estimation: For instance, in the DFT numerical study Ref.~\cite{velocity}, under different chosen magnetic order ansatzs, the sound speeds are found to differ by such a ratio.

The spin state with lower energy density can be viewed as the ``background'' domain, while the one with slightly higher energy density forms ``island'' domains over the background, which are considered as extended scatterers (their sizes thus must be larger than the phonon wavelength, and soon we will justify the self-consistency of this assumption). It is easy to calculate, given the sound speed ratio above, that the typical deflection angle between the in-coming and out-going sound waves off such an extended scatterer to satisfy $\cos\delta\theta\approx 0.99$.

Let $\ell_0$ be the phonon mean free path at large $B$, which is about $50\mu\mathrm{m}$ according to Ref.~\cite{Takagi2}. On the other hand, let $\ell$ be the phonon mean free path at the magnetic field of interest, and $n$ be the density of ``island'' domains at that magnetic field. If refraction is the main scattering mechanism as we supposed, then
\begin{eqnarray}
\frac{1}{\ell} \sim \frac{1}{\ell_0}+\sqrt{n} (1-\cos\delta\theta).
\end{eqnarray}
From the $\kappa$ comparison, we know that in regime of $B$ of interest (but not exactly at the energy density crossings), $\ell$ is about half of $\ell_0$. An ``island'' domain density of ${n}\sim 1\mu\mathrm{m}^{-2}$ would produce sufficient scattering; importantly, this is consistent with the assumptions we made -- the domain length scale bounded by $1/\sqrt{n}$ is indeed larger than the typical phonon wavelength which is below $0.1\mu\mathrm{m}$ for $T>0.5\K$. At the energy density crossings, ``background'' and ``islands'' exchange roles and domain wall become denser, further reducing $\kappa$ to form the observed dips.

\subsection{Heat Capacity}
\label{ssect_heatC}

In the above we made our proposal and showed it can reproduce the observed effects under mild assumptions. There are some important questions about our proposal remaining to be justified.

We assumed the distribution of the competing spin state domains to be able to thermally re-equilibriate as we change the temperature during the experiments, so that the temperature dependence of the data can be explained. This assumption is in contrast to the familiar case of magnetic domains in permanent magnets or the case of structural domains, which, without annealing, get frozen rather than re-equilibriate at low temperatures. This contrast is due to the difference in the energy barrier. In fact, it is very reasonable for spin domains to be able to thermally re-equilibriate -- for instance, it is well-known that most ferromagnetic materials cannot be used as permanent magnets, precisely because their magnetic domains can easily thermally re-equilibriate rather than remain frozen below the Curie temperature; those that can be used as permanent magnets are in fact special cases, in the sense that all of them have special material properties that prevent thermal re-equilibriation, see Appendix \ref{app_domain_dyn} for a review. Therefore, while we do not assume the detailed knowledge of the nature of our proposed competing spin states in the $8\T < B < 12\T$ regime, it is reasonable to assume that their distribution follows thermal equilibrium (or with insignificant hysteresis).

Then, the natural question to ask is: Why are the proposed energy crossings not leading to signatures in the heat capacity data, if the competing domains are not frozen but involved in the thermodynamics? Our task in the remaining of this section is to estimate this contribution to the heat capacity, and show the result is consistent with the currently heat capacity measurement data.

To estimate the heat capacity contribution from the competing domains, we may think of the situation at each 2D layer of $\RuCl$ as an  effective 2D Ising model (and the coupling between layers is negligible). In this picture, we think of each effective ``site'' as representing a small region of the material, which can be in either of the two competing spin states -- so such an effective ``site'' is not an actual lattice site in $\RuCl$. Let each effective ``site'' occupy an area $r^2$; while there is no precise rule to fix the value of $r^2$, it cannot be smaller than, say, some tens of unit cells, because it would be impossible to consider any notion of ``spin state'' in too small an area. The $hs^z_i$ term in the Ising model, then, represents the energy difference between the two competing spin states over an area of $r^2$, while the $Js^z_i s^z_j$ term in the Ising model represents the domain wall energy.

With this perspective, we can use some known results about the thermal properties of the Ising model. The energy crossings correspond to when $h$ changes sign while $T<T_c$ in the Ising model. It is known that the heat capacity rises by a factor of order $1$ as $h$ approaches $0$ from either side, and has only a small discontinuity near $h=0$, instead of any divergence or prominent increase. (If there is no hysteresis, there will only be a kink at $h=0$, and an insignificant hysteresis turns the kink into a small discontinuity near $h=0$.) In particular, when $T\ll T_c$, the rise of heat capacity near $h=0$ is by a factor of order $1$ compared to when $h$ is large, and as $T$ increases this factor becomes small. Thus, any heat capacity feature due to the proposed energy crossings should be most easily observed at low temperatures.

\begin{figure}
    \centering
    \includegraphics[width=0.49\textwidth]{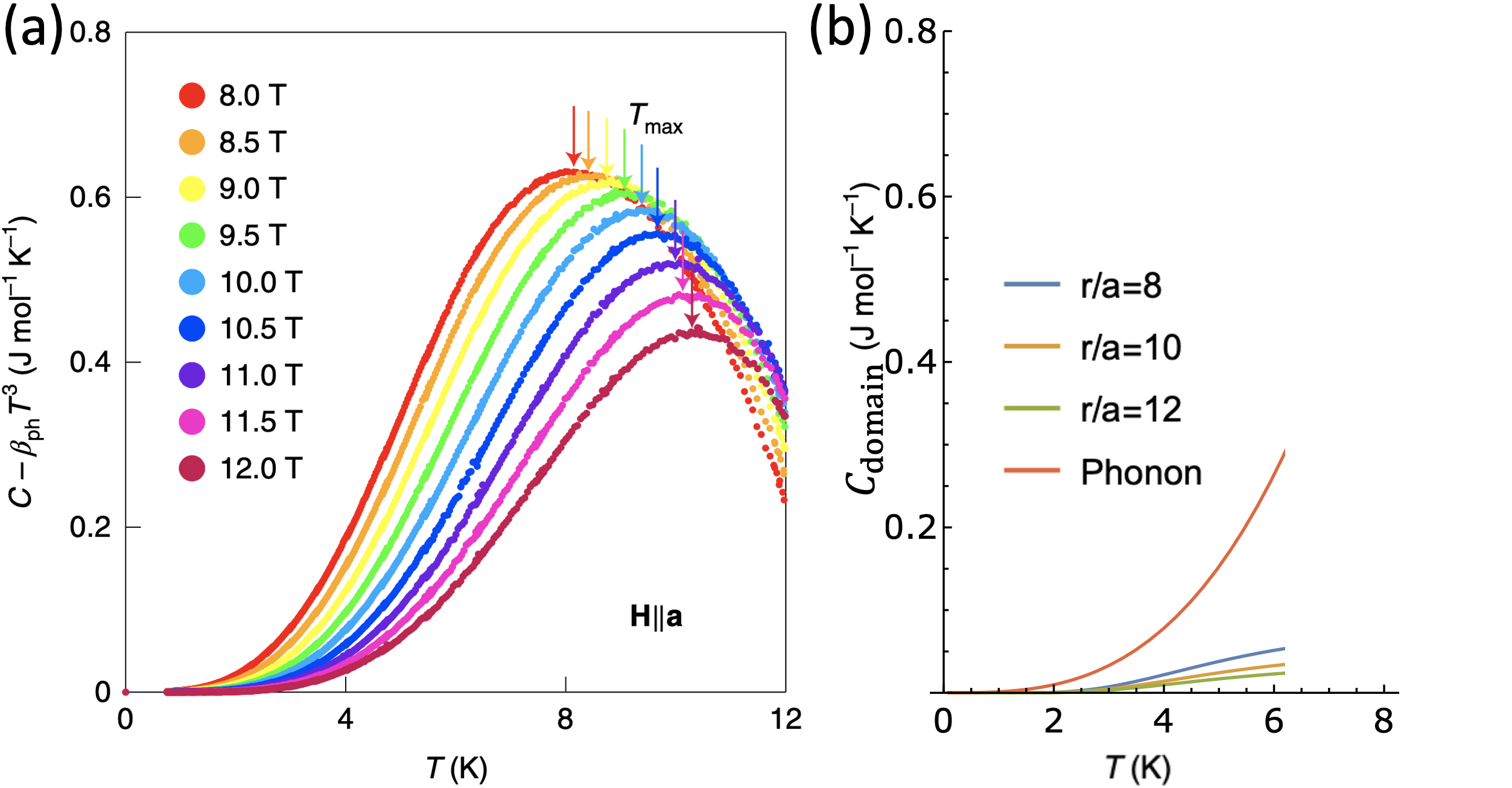}
    \caption{(a) Measured heat capacity in $\RuCl$ reprinted from \cite{heatcapacity}. The acoustic phonon contribution $\beta_{ph} T^3$ is subtracted, and the remaining contribution (plotted) is found to be exponentially small at low temperatures. (b) Our theoretical estimation \eqref{heat_capacity} (assuming $T_c\approx 10\mathrm{K}$), which is indeed much smaller than $\beta_{ph} T^3$ at low temperatures, and upper bounded by the data in (a) at all temperatures.}
    \label{fig:heat_capacity}
\end{figure}

Now, we shall show that at low temperatures, $C_{domain}\ll C_{phonon}$, so the order $1$ rises in $C_{domain}$ near the dips do not lead to any significant feature, in consistency with the experimental data. Since we are interested in $T\ll T_c$, it suffices to use mean field theory to estimate the scale of $C_{domain}$. The self-consistent equation for the mean field $\bar{s}$ is
\begin{eqnarray}
\bar{s}=\tanh(\bar{s}T_c/T).
\end{eqnarray}
As we are well away from $T_c$, we have $\bar{s} \sim 1$ (this is compatible with $n\sim 1\mu\mathrm{m}^{-2}$ estimated at the end of Section \ref{ssect_domains}) and $d\bar{s}/dT\sim -4(T_c/T^2) e^{-2T_c/T}$. Then we use $C_{domain}=(1/V)(dE/dT)$ and $E/V \sim (k_B T_c/2)\bar{s}^2$ to find the average heat capacity per unit cell:
\begin{eqnarray}
\label{heat_capacity}
C_{domain} a^3 \sim 4k_B\frac{a^2}{r^2}\frac{T_c^2}{T^2}\: e^{-2T_c/T}.
\end{eqnarray}

Now we compare this result with experiments. In Ref.~\cite{heatcapacity}, the heat capacity of $\RuCl$ is measured. The contribution  $C_{ph}a^3/\mathrm{mol}=\beta_{ph} T^3$ from acoustic phonons is extracted to find $\beta_{ph}\approx 1.22\, \rm{mJ\, mol^{-1} K^{-4}}$, and after subtracting off the phonon contribution, the remaining heat capacity is reproduced in Fig. \ref{fig:heat_capacity}(a). This remaining heat capacity should be the contributions from the spin system, and is found to be exponentially small at low temperatures, with a characteristic scale of about $10\mathrm{K}$. Our $C_{domain}$ contribution should be upper bounded by this measured remaining part ($C_{domain}$ can be smaller than total spin system contribution because the spin system may also contain gapped magnon contributions, etc.); moreover, it should be much smaller than the phonon contribution at low $T$ as mentioned before. In Fig. \ref{fig:heat_capacity}(b), we plotted $C_{domain}$ estimated using \eqref{heat_capacity}, assuming $T_c \approx 10\mathrm{K}$ and $r\sim 10a$. Indeed, we find $C_{domain}\ll C_{phonon}$ at small $T$, and is also smaller than the measured total spin system contribution. (We are only plotting up to $T\leq 6\mathrm{K}$ because we only need the result for $T\ll T_c$, and moreover the mean field estimation we employed is indeed only suitable for this regime.)

\section{Relation to Thermal Hall}
\label{sect_Hall}

As discussed in the introduction, another notable thermal transport phenomenon is the surprisingly large thermal Hall effect, which starts to ramp up at around $4\K$ and peaks at around $10\K$ \cite{Ong2}. While initially speculated to have an origin in the chiral Majorana edge mode of Kitaev spin liquid, later phenomenological theory suggests an origin in gapped magnons with topological Chern band \cite{Kimmagnon}, and the predictions of this theory agree well with the recent data \cite{Ong2}.

It is natural to ask whether the interesting phenomena in the longitudinal thermal transport and the thermal Hall transport are related. The temperature regimes of the two phenomena are different: as the temperature rises to where $\kappa_H$ starts to ramp up, the dips in the longitudinal $\kappa$ are already diminishing. The plausible heat carriers are also different: as we have argued in Section \ref{sect_general}, at the low temperatures at which the dips in $\kappa$ are seen, the heat current is carried by acoustic phonons; on the other hand, the thermal Hall current are carried by gapped magnons, hence only occurring at the temperature scale of the gap, $\sim 10\K$. However, the plausible mechanisms behind these two phenomena do have an important connection -- they both involve low energy spin states competing in the regime $8\T < B < 12\T$.

Ref.~\cite{Kimmagnon} conducted a mean field study of the Heisenberg+Kitaev model of $\RuCl$ in the regime of $8\T < B < 12\T$. While \emph{a priori} the mean field approximation is not controlled in such a strongly coupled and strongly frustrated system, the reproduction of the observed thermal Hall curve from the resulting topological magnon band suggests that it indeed captured some essential physics. (Remarkably, neutron scattering \cite{magneticorder2, magneticorder} shows no obvious spectral line in this regime, which means the proposed magnons should be strongly interacting and are not long-lived quasiparticles, but their topological Chern band physics retains.) The mean field analysis finds multiple competing low energy spin states, whose energy densities cross each other with the tuning of $B$, in a manner that is more complicated than the schematic Fig.~\ref{fig:energy_crossings}.

While our phenomenology study does not rely on the detailed knowledge of the competing states, Ref.~\cite{Kimmagnon} contained some detailed mean field solutions. They are all spin ordered states, with large spin ordered unit cells -- some tens of the original unit cell -- due to the strong frustration. These are some reasonable candidates for the possible nature of the competing spin states, and they are compatible with our assumption that $r^2$ is at least some tens of $a^2$ when discussing the domain contribution to heat capacity. (It is also possible that in the actual material, the low energy spin states are not ordered, which if true would be a more interesting situation, but this is subjected to further experimental studies.)

From different starting points and via different approaches, the analysis of \cite{Kimmagnon} and our analysis both point to the presence of competing spin states in the $8\T < B < 12\T$ regime, in a compatible manner, and together the analyses explain and reproduce the seemingly puzzling thermal transport phenomenon observed in $\RuCl$.

\section{Summary and Predictions}
\label{sect_predictions}

In this paper we provided phenomenological explanations for the puzzling ``oscillatory'' behavior in the longitudinal thermal conductivity of $\RuCl$. The dips in the thermal conductivity are caused by two mechanisms in different regimes of magnetic fields. Critical scattering causes the dips at small magnetic fields, as analyzed in Section \ref{sect_firstset}, in agreement with \cite{Takagi1}. At higher magnetic fields in the debated ``quantum spin liquid'' regime, we propose in Section \ref{sect_secondset} that there are competing spin states forming domains, and the dips in the thermal conductivity are caused by domain wall scattering. In this mechanism we do not have to assume detailed knowledge about the nature of these competing states -- which we believe cannot be determined by the currently available experimental data. Our simple proposal can explain the significant as well as the subtle features in the data, and is moreover consistent with other measured properties of the material. Our proposal of competing spin states is also consistent with the theory for thermal Hall conductivity \cite{Kimmagnon} (Section \ref{sect_Hall}), hence forming a coherent phenomenological picture for what is happening in the thermal transport in $\RuCl$.

We make some predictions regarding our proposal of competing spin states in the regime $8\T < B < 12\T$, to be examined by future experiments. In our proposal the domain wall scattering is primarily due to refraction, which requires different acoustic velocities in the two competing spin states (in agreement with \cite{velocity}). To examine this difference, one can go to the cold enough temperature so that the system is almost entirely in the lowest energy spin state. For $B$ on the two sides of an energy crossing point (a dip in $\kappa$), the lowest energy spin states are different. Then their difference in acoustic velocity will manifest in sound propagation, thermal conductivity ($\propto v^{-2}$) and heat capacity ($\propto v^{-3}$). A measurement precision to about $10\%$ is needed.

\begin{acknowledgments}
We are grateful to Steven~Kivelson, Zheng~Liu and Patrick~Lee for valuable discussions, and we acknowledge Chuan~Chen for involvement at an early stage of this project. We thank the authors of the experimental papers \cite{Ong1, Ong2, Takagi1, Matsuda1, heatcapacity} for authorizing us to use the data and plots reprinted from these papers. Z.-D.~F. and J.-Y.~C. are supported by NSFC under Grants No.~12174213. X.-Q.~S is supported by the Gordon and Betty Moore Foundation's EPiQS Initiative through Grant GBMF8691, and by the Deutsche Forschungsgemeinschaft (DFG) under Germany's Excellence Strategy (EXC2111-390814868).
\end{acknowledgments}

\appendix

\section{Some Derivations in the Theory of Critical Scattering}
\label{app_crit_scatt}

In this appendix we derive \eqref{ImChi_factors}. We do so by performing two computations for $\chi(\omega, q)$, one hydrodynamical, and one quantum mechanical, and matching the results. First, the response function $\chi$, as the name implies, is the physical measurable quantity in linear response, therefore we expect to describe it by some phenomenological computation in hydrodynamics. Here, the perturbation is an energy density operator that generates a temperature gradient \cite{Luttinger}, and the measured quantity is also the energy density. We set up a perturbation $\Delta T(t, \mathbf{r})=T(t, \mathbf{r})-T_{eq}$; since we need to know the $\omega, q$ dependence in $\chi$, we need a non-uniform, time-dependent $\Delta T$. It suffices to consider a quenched setting
\begin{eqnarray}
\Delta T(t, \mathbf{r})/T = - \epsilon \, \theta(-t) e^{i\mathbf{q\cdot r}}.
\end{eqnarray}
Since for $t<0$ the perturbation is time-independent, the energy density at $t\leq 0$ is, by definition, $j^0(t\leq 0, \mathbf{r})=\chi(\omega=0, \mathbf{q}) e^{i\mathbf{q\cdot r}}\, \epsilon$. The behavior at $t>0$ can be derived by the diffusion equation and the conservation equation:
\begin{eqnarray}
j^i =-D \partial_i j^0, \ \ \ \partial_t j^0= - \partial_i j^i,
\end{eqnarray}
which lead to
\begin{eqnarray}
&&j^0(t, \mathbf{r})=j^0(t=0, \mathbf{r})\, e^{-Dq^2 t},  \nonumber \\[.2cm]  
&&\int_0^\infty dt \: j^0(t, \mathbf{r}) \: e^{i\omega t} 
=\frac{\chi(0, \mathbf{q})}{Dq^2-i\omega} \, e^{i\mathbf{q\cdot r}} \, \epsilon \ .
\end{eqnarray}
On the other hand, we can also compute the result in the general quantum mechanical linear response theory:
\begin{eqnarray}
&&\int_0^\infty dt \: j^0(t, \mathbf{r}) \: e^{i\omega t} \\[.2cm] 
=&&\int_0^\infty dt \: e^{i\omega t} \int_{-\infty}^0 dt' \: \chi(t-t', \mathbf{q}) \, e^{i\mathbf{q\cdot r}} \epsilon \nonumber \\[.2cm]
=&& \int_0^\infty dt \: e^{i\omega t} \int_{-\infty}^0 dt' \int_{-\infty}^\infty \frac{d\omega'}{2\pi} e^{-i\omega'(t-t')} \chi(\omega', \mathbf{q}) \, e^{i\mathbf{q\cdot r}} \epsilon \nonumber \\[.2cm]
=&& \int_0^\infty dt\, \int_{-\infty}^\infty \frac{d\omega'}{2\pi i} \, \frac{e^{i(\omega-\omega')t}}{\omega'-i0^+} \:\chi(\omega', \mathbf{q}) \, e^{i\mathbf{q\cdot r}} \epsilon \nonumber \\[.2cm]
=&& \int_{-\infty}^\infty \frac{d\omega'}{2\pi} \, \frac{\chi(\omega', \mathbf{q}) \, e^{i\mathbf{q\cdot r}} \epsilon}{(\omega'-i0^+)(\omega-\omega'+i0^+)}.
\end{eqnarray}
Using the analytical property of $\chi(\omega', \mathbf{q})$, the above can be evaluated as contour integral and we find
\begin{eqnarray}
&& \frac{i}{\omega} \, \left[ \chi(0, \mathbf{q})-\chi(\omega, \mathbf{q}) \right] \,  \, e^{i\mathbf{q\cdot r}} \epsilon \ .
\end{eqnarray}
Equating the results from the hydrodynamical computation and the general quantum mechanical computation, we have
\begin{eqnarray}
\chi(\omega, \mathbf{q})= \chi(0, \mathbf{q}) \left[ 1 - \frac{\omega}{\omega+iDq^2} \right] \ .
\end{eqnarray}
Taking the imaginary part and using the analytical property that $\chi(0, \mathbf{q})$ is real leads to \eqref{ImChi_factors}.

\section{The Mechanism of Low Energy Dynamical Defect Scatterers}
\label{app_dyn_defect}

In this appendix, we give another mechanism in which thermal conductivity $\kappa$ can have dips with varying external magnetic field. If there is some kind of dynamical defects in the material coupled to the phonons, $\kappa$ decrease when the typical energy transition of the dynamical defects is near the typical phonon energy; if the energy levels of the defects vary with the magnetic field, dips in $\kappa$ will develop with varying magnetic field. Here we consider a two level dynamical defect derivation to illustrate the idea. This mechanism has appeared in the appendix of \cite{defect4}. We will explain why we believe this mechanism is not applicable to the observation in $\RuCl$.

Suppose the energy separation of a two-level dynamical defect is $\Delta$, so that in the defect's diagonal basis we have $H_{defect}=-(\Delta/2)\sigma^3$. Meanwhile, consider the coupling between phonons and the defect:
\begin{eqnarray}
H_{int}&&=g (\partial\cdot u) \sigma^1   \\ \nonumber
&&\sim \frac{g \hbar}{\sqrt{2\rho V}} \sum_k \sqrt{\frac{1}{\hbar \omega_k}} ik (a_k + a_{-k}^\dagger ) \sigma^1
\end{eqnarray}
The second line is schematic, as we have neglected details such as the phonon polarizations. (There can also be $\sigma^3$ couplings but they are unimportant for the development of dips.) Then we compute the T-matrix, which give us the cross-section of phonon scattering off the defect:
\begin{eqnarray}
T_{k'k}^{\alpha} = \langle k',\alpha| H_{int} G(E_{in}+i0^+) H_{int} | k,\alpha \rangle
\end{eqnarray}
where $\alpha=-$ and $+$ represent the two levels of the defect, and $G(E)=\frac{1}{E-H_{defect}}$ is the defect Green's function. The T-matrix gives the collision term in Boltzmann equation of phonon:
\begin{eqnarray}
\vec{v_k}\cdot \frac{\partial f_0}{\partial T} \vec{\nabla} T &&= -\left.\frac{df}{dt}\right|_{collison}
\end{eqnarray}
where $df/dt|_{collison}=g(k)/\tau(k)$, with $g(k)$ = $f(k)-f_0(k)$ and
\begin{eqnarray}
\frac{1}{\tau(k)} 
=\frac{VN_d }{\pi v\hbar^2} k^2 
 \left[|T_{k'k}^-|^2 \frac{1}{1+e^{-\frac{\Delta}{T}}} +|T_{k'k}^+|^2 \frac{1}{1+e^{\frac{\Delta}{T}}} \right] \nonumber \\
 + \frac{1}{\tau_0} \hspace{6cm}
\label{defect_tau}
\end{eqnarray}
in which $\tau_0$ is the relaxation time contributed by boundary scattering, $N_d$ is the total number of defects, and $v$ is acoustic velocity.

The matrix element $|T_{k'k}^{\alpha}|^2$ can be computed:
\begin{eqnarray}
|T_{k'k}^{\alpha}|^2 = \frac{g^4 \hbar^2 k^2}{4\rho^2 V^2 v^2} \frac{1}{(\hbar \omega_k -\Delta)^2+ \frac{\Gamma_\alpha^2}{4}}
\end{eqnarray}
where $\Gamma_\alpha = \frac{1}{4\pi} \frac{g^2 \Delta^3}{\rho v^5 \hbar^3} f_\alpha$, and $f_\alpha = 1+\frac{1}{e^{\Delta/T}-1}$ and $\frac{1}{e^{\Delta/T}-1}$ when $\alpha=-$ and $+$ respectively. Substituting into the Boltzmann equation and solving for the thermal current response, one can find the thermal conductivity to be:
\begin{eqnarray}
\kappa=\frac{k_B^4 T^3}{2\pi^2 \hbar^3 v}\tau_0 F
\end{eqnarray}
where $F$ is the dimensionless integral
\begin{eqnarray}
F=\int_0^{\frac{T_D}{T}}dx \frac{x^4e^x}{(e^x-1)^2} \frac{\tau(x)}{\tau_0}
\end{eqnarray}
in which $x$ is the dimensionless variable $\hbar\omega_k/k_BT$. 

\begin{figure}
    \centering
    \includegraphics[width=0.35\textwidth]{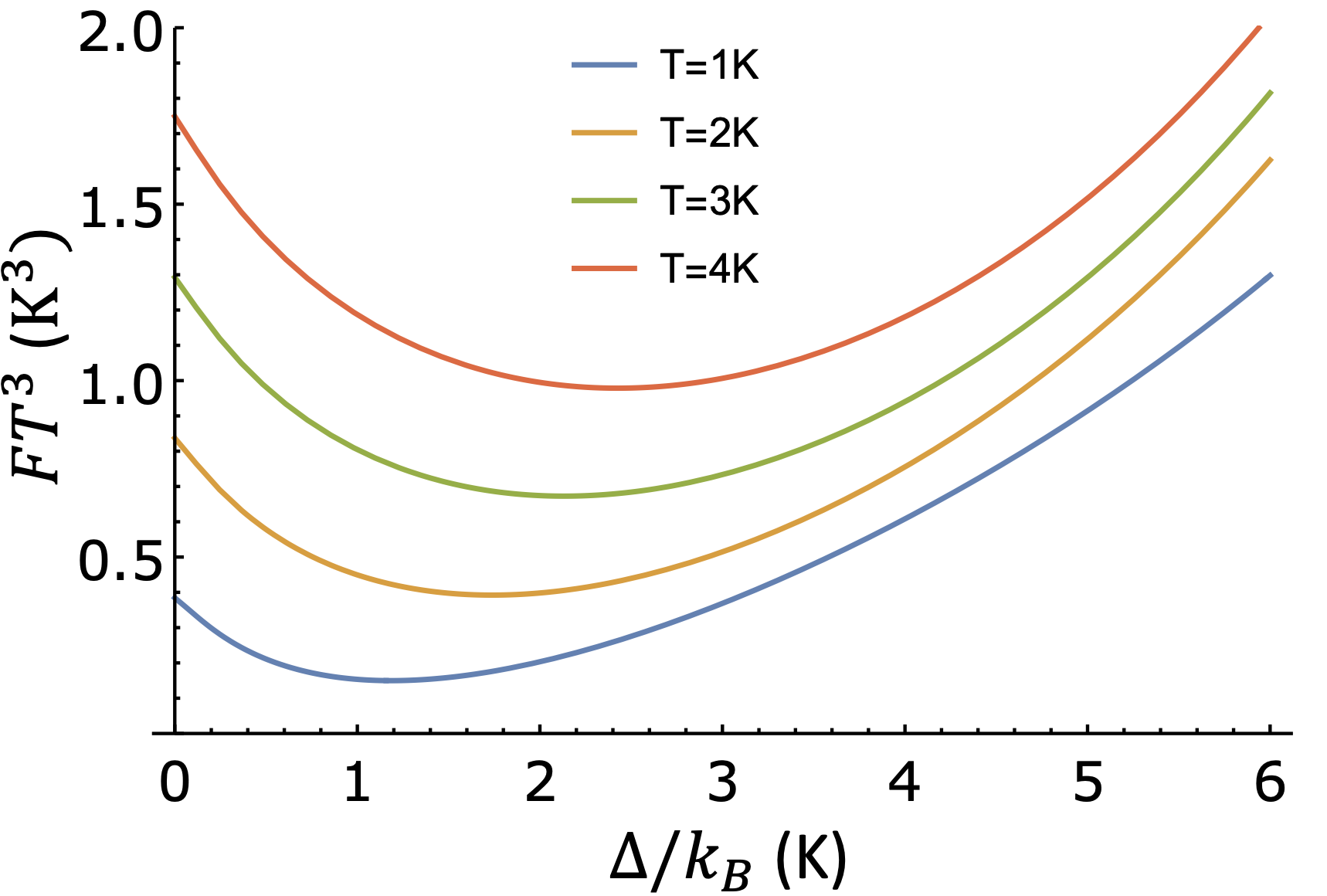}
    \caption{$FT^3 (\propto \kappa)$ varying with $\Delta$. Unfortunately the minimum moves with $T$.}
    \label{fig:kappa}
\end{figure}

Now, if we consider $\Delta$ as the tuning parameter that varies with the magnetic field, we can find that $\kappa$ develops a minimum. We plot this in Fig. \ref{fig:kappa}, using parameters suitable for $\RuCl$: $v=600\, \rm{m/s}$, $\rho=2000 \, \rm{kg/m^3}$, $\tau_0=50\mu m/v\sim 10^{-7}\, \rm{s}$, and assuming $N_d/V=10^{-3}\, \rm{nm^{-3}}$ and $g=0.1\, \rm{eV}$. The minimum is produced because in $F$, the $\tau(x)/\tau_0$ factor is a function that is almost constantly $1$ for most $x$, except for a region where the function drops to almost $0$; the width of this region is determined by when the first term in \eqref{defect_tau} becomes larger than $1/\tau_0$, and physically this region of $x$ presents the energies of those phonons whose $\omega_k$ are close enough to $\Delta$ so that the scatterings with the dynamical defects dominate over the boundary scatterings. (Thus, the width of this suppressed region in $\tau$ is \emph{not} controlled by $\Gamma$, although $\Gamma$ is usually called the ``width''. We can even set $\Gamma\rightarrow 0$ here and the result remains unchanged.) When this region overlaps with the peak of the $x^4e^x/(e^x-1)^2$ (which represents the majority of the phonon excitations) in $F$, the integral will be suppressed. 

The problem of applying this mechanism to explain the dips of $\kappa$ in $\RuCl$ is that the positions of the dips vary with temperature, but in the experimental data, the positions of the dips barely move. (Another problem is it is hard to produce multiple nearby dips, because the minima due to two $\Delta$'s will usually combine into one broad minimum.)

\section{Domain Configurations: Re-Equilibriate versus Frozen}
\label{app_domain_dyn}

In this section, we review the physics of whether domains tend to re-equilibriate or to freeze at low temperatures \cite{domainconfig}. In some familiar examples such as permanent magnets and structural domains, the domains are frozen at low temperatures, but in others they do not. Four kinds of ``energies'' are involved in the competition: The energy cost localized on domain walls (such as the exchange energy in ferromagnet), the long range energy cost between domains (such as the electromagnetic energy in ferromagnet), the entropy contribution in the free energy, and the ``coercive energy'' against moving the domain walls. In all cases the local domain wall energy and the entropy are present, so what distinguish the cases are the long range energy and the ``coercive energy''. What determines whether domains re-equilibriate or freeze is whether the ``coercive energy'' is minor or significant. In the below we briefly review the different cases.

In the familiar Ising model, there is neither a long range energy nor a ``coercive energy''. Domains easily re-equilibriate. The magnetization has a finite jump when $B$ crosses $0$ and there is no hysteresis.

In ferromagnets like Fe/Co/Ni, there is the long range electromagnetic energy between spin domains and no significant ``coercive energy''. Again domains easily re-equilibriate. But the difference with the Ising model is that, here the magnetization is ``softer'': Even as $T\rightarrow 0$, the magnetization changes continuously (though rapidly) as $B$ crosses $0$, but there is no finite jump. This is because at large distances, the long range electromagnetic energy cost, which prefers domains to be small, dominates over the local domain wall energy, which prefers domains to be large; hence at $B=0$ the magnetization must be $0$.

In ferromagnets that can be used as permanant magnets, there is still the long range electromagnetic energy between spin domains, but the ``coercive energy'' is significant, making the magnetization ``hard''. These materials are usually alloy with a lot of disorders, and the energy costs for spin domain walls to move across certain kinds of disorders are very high, so that the domains tend to freeze rather than re-equilibriate, leading to significant hysteresis. (Among ferromagnetic materials, the coercivity between different materials can differ by a factor of $10^{7}$ \cite{domainconfig}.) The long range electromagnetic energy does not play a crucial role here, therefore, in other kinds of highly ``coercive'' situations such as structural domains where there is no long range energy involved, the qualitative behavior of frozen domain walls would remain similar.

For our proposal of competing spin states in $\RuCl$ in $8\T <B<12\T$, there is no obvious element to produce a significant coercive effect, therefore it is a reasonable assumption that the spin states tend to re-equilibriate. Whether there is a long range energy between the domains would depend on more details of the nature of the spin states.

\bibliographystyle{apsrev4-2}
\bibliography{refs}

@article{5/2quantumhall,
  title={Observation of half-integer thermal Hall conductance},
  author={Banerjee, Mitali and Heiblum, Moty and Umansky, Vladimir and Feldman, Dima E and Oreg, Yuval and Stern, Ady},
  journal={Nature},
  volume={559},
  number={7713},
  pages={205--210},
  year={2018},
  publisher={Nature Publishing Group UK London},
  doi={10.1038/s41586-018-0184-1}
}

@article{kitaevspinliquid,
  title={Anyons in an exactly solved model and beyond},
  author={Kitaev, Alexei},
  journal={Annals of Physics},
  volume={321},
  number={1},
  pages={2--111},
  year={2006},
  publisher={Elsevier},
  doi={10.1016/j.aop.2005.10.005}
}

@article{Ong1,
  title={Oscillations of the thermal conductivity in the spin-liquid state of $\alpha$-RuCl3},
  author={Czajka, Peter and Gao, Tong and Hirschberger, Max and Lampen-Kelley, Paula and Banerjee, Arnab and Yan, Jiaqiang and Mandrus, David G and Nagler, Stephen E and Ong, NP},
  journal={Nature Physics},
  volume={17},
  number={8},
  pages={915--919},
  year={2021},
  publisher={Nature Publishing Group UK London},
  doi={10.1038/s41567-021-01243-x}
}

@article{Ong2,
  title={Planar thermal Hall effect of topological bosons in the Kitaev magnet $\alpha$-RuCl3},
  author={Czajka, Peter and Gao, Tong and Hirschberger, Max and Lampen-Kelley, Paula and Banerjee, Arnab and Quirk, Nicholas and Mandrus, David G and Nagler, Stephen E and Ong, N Phuan},
  journal={Nature Materials},
  volume={22},
  number={1},
  pages={36--41},
  year={2023},
  publisher={Nature Publishing Group UK London},
  doi={10.1038/s41563-022-01397-w}
}

@article{Matsuda1,
  title={Evidence for a Phase Transition in the Quantum Spin Liquid State of a Kitaev Candidate $\alpha$-RuCl3},
  author={Shota Suetsugu and Yuzuki Ukai and Masaki Shimomura and Masashi Kamimura and Tomoya Asaba and Yuichi Kasahara and Nobuyuki Kurita and Hidekazu Tanaka and Takasada Shibauchi and Joji Nasu and Yukitoshi Motome and Yuji Matsuda},
  journal={Journal of the Physical Society of Japan},
  number={12},
  pages={124703},
  year={2022},
  publisher={The Physical Society of Japan},
  doi={10.7566/JPSJ.91.124703}
}

@article{Matsuda2,
  title={Majorana quantization and half-integer thermal quantum Hall effect in a Kitaev spin liquid},
  author={Kasahara, Y and Ohnishi, T and Mizukami, Y and Tanaka, O and Ma, Sixiao and Sugii, K and Kurita, N and Tanaka, H and Nasu, J and Motome, Y and Shibauchi, T and Matsuda, Y},
  journal={Nature},
  volume={559},
  number={7713},
  pages={227--231},
  year={2018},
  publisher={Nature Publishing Group UK London},
  doi={10.1038/s41586-018-0274-0}
}

@article{Takagi1,
  title={Origin of oscillatory structures in the magnetothermal conductivity of the putative Kitaev magnet $\alpha$-RuCl3},
  author={Bruin, JAN and Claus, RR and Matsumoto, Y and Nuss, J and Laha, S and Lotsch, BV and Kurita, N and Tanaka, H and Takagi, H},
  journal={APL Materials},
  volume={10},
  number={9},
  pages={090703},
  year={2022},
  publisher={AIP Publishing LLC},
  doi={10.1063/5.0101377}
}

@article{Takagi2,
  title={Robustness of the thermal Hall effect close to half-quantization in $\alpha$-RuCl3},
  author={Bruin, JAN and Claus, RR and Matsumoto, Y and Kurita, N and Tanaka, H and Takagi, H},
  journal={Nature Physics},
  volume={18},
  number={4},
  pages={401--405},
  year={2022},
  publisher={Nature Publishing Group UK London},
  doi={10.1038/s41567-021-01501-y}
}

@article{magneticorder,
  title = {Field-induced intermediate ordered phase and anisotropic interlayer interactions in $\ensuremath{\alpha}\text{\ensuremath{-}}{\mathrm{RuCl}}_{3}$},
  author = {Balz, C. and Janssen, L. and Lampen-Kelley, P. and Banerjee, A. and Liu, Y. H. and Yan, J.-Q. and Mandrus, D. G. and Vojta, M. and Nagler, S. E.},
  journal = {Phys. Rev. B},
  volume = {103},
  issue = {17},
  pages = {174417},
  numpages = {12},
  year = {2021},
  month = {May},
  publisher = {American Physical Society},
  doi = {10.1103/PhysRevB.103.174417},
  url = {https://link.aps.org/doi/10.1103/PhysRevB.103.174417}
}

@article{magneticorder2,
  title = {Finite field regime for a quantum spin liquid in $\ensuremath{\alpha}\text{\ensuremath{-}}{\mathrm{RuCl}}_{3}$},
  author = {Balz, Christian and Lampen-Kelley, Paula and Banerjee, Arnab and Yan, Jiaqiang and Lu, Zhilun and Hu, Xinzhe and Yadav, Swapnil M. and Takano, Yasu and Liu, Yaohua and Tennant, D. Alan and Lumsden, Mark D. and Mandrus, David and Nagler, Stephen E.},
  journal = {Phys. Rev. B},
  volume = {100},
  issue = {6},
  pages = {060405},
  numpages = {6},
  year = {2019},
  month = {Aug},
  publisher = {American Physical Society},
  doi = {10.1103/PhysRevB.100.060405},
  url = {https://link.aps.org/doi/10.1103/PhysRevB.100.060405}
}

@article{heatcapacity,
  title={Thermodynamic evidence for a field-angle-dependent Majorana gap in a Kitaev spin liquid},
  author={Tanaka, O and Mizukami, Y and Harasawa, R and Hashimoto, Kenichiro and Hwang, K and Kurita, N and Tanaka, H and Fujimoto, S and Matsuda, Y and Moon, E-G and Shibauchi, T},
  journal={Nature Physics},
  volume={18},
  number={4},
  pages={429--435},
  year={2022},
  publisher={Nature Publishing Group UK London},
  doi={10.1038/s41567-021-01488-6}
}

@article{abinitio,
  title={Crystal structure and magnetism in $\alpha$- RuCl 3: An ab initio study},
  author={Kim, Heung-Sik and Kee, Hae-Young},
  journal={Physical Review B},
  volume={93},
  number={15},
  pages={155143},
  year={2016},
  publisher={APS},
  doi={10.1103/PhysRevB.93.155143}
}

@article{velocity,
  title={Acoustic phonon dispersion of $\alpha$- RuCl 3},
  author={Lebert, Blair W and Kim, Subin and Prishchenko, Danil A and Tsirlin, Alexander A and Said, Ayman H and Alatas, Ahmet and Kim, Young-June},
  journal={Physical Review B},
  volume={106},
  number={4},
  pages={L041102},
  year={2022},
  publisher={APS},
  doi={10.1103/PhysRevB.106.L041102}
}

@article{Kimmagnon,
  title = {Topological magnons for thermal Hall transport in frustrated magnets with bond-dependent interactions},
  author = {Zhang, Emily Z. and Chern, Li Ern and Kim, Yong Baek},
  journal = {Phys. Rev. B},
  volume = {103},
  issue = {17},
  pages = {174402},
  numpages = {9},
  year = {2021},
  month = {May},
  publisher = {American Physical Society},
  doi = {10.1103/PhysRevB.103.174402},
  url = {https://link.aps.org/doi/10.1103/PhysRevB.103.174402}
}

@article{Luttinger,
  title={Theory of thermal transport coefficients},
  author={Luttinger, JM},
  journal={Physical Review},
  volume={135},
  number={6A},
  pages={A1505},
  year={1964},
  publisher={APS},
  doi={10.1103/PhysRev.135.A1505}
}

@article{Stern,
  title={Thermal conductivity at the magnetic transition},
  author={Stern, Harry},
  journal={Journal of Physics and Chemistry of Solids},
  volume={26},
  number={1},
  pages={153--161},
  year={1965},
  publisher={Elsevier},
  doi={10.1016/0022-3697(65)90082-X}
}

@ARTICLE{defect1,
       author = {{Phillips}, W.~A.},
        title = "{Tunneling states in amorphous solids}",
      journal = {Journal of Low Temperature Physics},
         year = 1972,
        month = may,
       volume = {7},
       number = {3-4},
        pages = {351-360},
          doi = {10.1007/BF00660072},
       adsurl = {https://ui.adsabs.harvard.edu/abs/1972JLTP....7..351P}
}

@article{defect2,
  title = {Large extrinsic phonon thermal Hall effect from resonant scattering},
  author = {Sun, Xiao-Qi and Chen, Jing-Yuan and Kivelson, Steven A.},
  journal = {Phys. Rev. B},
  volume = {106},
  issue = {14},
  pages = {144111},
  numpages = {15},
  year = {2022},
  month = {Oct},
  publisher = {American Physical Society},
  doi = {10.1103/PhysRevB.106.144111},
  url = {https://link.aps.org/doi/10.1103/PhysRevB.106.144111}
}

@article{defect3,
author = {Haoyu Guo  and Darshan G. Joshi  and Subir Sachdev },
title = {Resonant thermal Hall effect of phonons coupled to dynamical defects},
journal = {Proceedings of the National Academy of Sciences},
volume = {119},
number = {46},
pages = {e2215141119},
year = {2022},
doi = {10.1073/pnas.2215141119},
URL = {https://www.pnas.org/doi/abs/10.1073/pnas.2215141119}
}

@article{defect4,
  title = {Origin of the Phonon Hall Effect in Rare-Earth Garnets},
  author = {Mori, Michiyasu and Spencer-Smith, Alexander and Sushkov, Oleg P. and Maekawa, Sadamichi},
  journal = {Phys. Rev. Lett.},
  volume = {113},
  issue = {26},
  pages = {265901},
  numpages = {5},
  year = {2014},
  month = {Dec},
  publisher = {American Physical Society},
  doi = {10.1103/PhysRevLett.113.265901},
  url = {https://link.aps.org/doi/10.1103/PhysRevLett.113.265901}
}

@ARTICLE{spinliquid1,
       author = {{Takagi}, Hidenori and {Takayama}, Tomohiro and {Jackeli}, George and {Khaliullin}, Giniyat and {Nagler}, Stephen E.},
        title = "{Concept and realization of Kitaev quantum spin liquids}",
      journal = {Nature Reviews Physics},
         year = 2019,
        month = mar,
       volume = {1},
       number = {4},
        pages = {264-280},
          doi = {10.1038/s42254-019-0038-2},
 primaryClass = {cond-mat.str-el},
}

@article{spinliquid2,
author = {C. Broholm  and R. J. Cava  and S. A. Kivelson  and D. G. Nocera  and M. R. Norman  and T. Senthil },
title = {Quantum spin liquids},
journal = {Science},
volume = {367},
number = {6475},
pages = {eaay0668},
year = {2020},
doi = {10.1126/science.aay0668}
}

@article{spinliquid3,
  title = {Quantum spin liquid states},
  author = {Zhou, Yi and Kanoda, Kazushi and Ng, Tai-Kai},
  journal = {Rev. Mod. Phys.},
  volume = {89},
  issue = {2},
  pages = {025003},
  numpages = {50},
  year = {2017},
  month = {Apr},
  publisher = {American Physical Society},
  doi = {10.1103/RevModPhys.89.025003},
  url = {https://link.aps.org/doi/10.1103/RevModPhys.89.025003}
}

@article{spinliquid4,
    author = "Savary, Lucile and Balents, Leon",
    title = "{Quantum spin liquids: a review}",
    primaryClass = "cond-mat.str-el",
    doi = "10.1088/0034-4885/80/1/016502",
    journal = "Rept. Prog. Phys.",
    volume = "80",
    number = "1",
    pages = "016502",
    year = "2017"
}

@book{domainconfig,
  title={Introduction to solid state physics},
  author={Kittel, Charles},
  year={2005},
  publisher={John Wiley \& sons, inc}
}

\end{document}